\documentclass[aps,showpacs,prl,toolkits,nofootinbib,preprint]{revtex4}
\usepackage{mathrsfs}
\usepackage{amsmath}
\usepackage{graphicx}
\usepackage{color}
\usepackage{epsfig}
\usepackage{subfigure}

\begin{document}

\title{Some Comments on Possible Preferred Directions
                   for the SETI Search}

\author{Shmuel Nussinov}

\email{nussinov@post.tau.ac.il} \affiliation{Tel Aviv University,
Sackler School Faculty of Sciences, \\Ramat Aviv, Tel Aviv 69978,
Israel, and \\Chapman University, Schmid College of Science,\\ Orange, California 92866, USA}

\begin{abstract}

The search for extraterrestrial intelligence by looking for signals
    from advanced technological civilizations has been ongoing for some
    decades. We suggest that it could possibly be made more efficient
    by focusing on stars from which the solar system can be observed
    via mini-eclipsings of the Sun by transiting planets.

\end{abstract}

%\pacs{ }

\maketitle

\vspace{.25in}

\noindent{\bf Introduction}

 The quest for extraterrestrial life and intelligence is as old as civilization itself. Yet, only recently (thanks to pioneering works of Shklovski, Sagan and others \cite{ShklovskiSagan} have serious searches like SETI been initiated. Sifting through all radio signals from the direction of ``favorable", nearby stars, these searches for signals from intelligent civilizations on planetary system around these stars.

 Our galaxy contains $\sim 10^{10}$ candidate solar type stars and the broadcasted frequency and method used for communicating information are not known. Finding some ``intelligent" signal in the vast inflow of data is challenging.

 Various ``guesses" of the likely wavelength used or the content of first communication have been made and can  be of crucial importance. \cite{DrakeSaganGott} The basic "Copernicean" tenet of Gott presented in
 Ref. \cite{DrakeSaganGott} and some estimates of Bracewell \cite{Bracewell} as to the total number of habitable planets and civilizations suggests that the galactic disc may provide preferred directions for the SETI search. Some of the potential Intelligent Technological Societies (ITS's) around the many stars there may be sufficiently advanced allowing it to generate very strong and very beamed broadcasting---compensating the geometric, $1/(R^2)$ decrease.

 Here we comment on a possible different guide as to preferred search directions motivated by the recent advances in searches of extrasolar planetary systems.  Since ITS's are more likely to beam towards habitable
planets.\footnote{Isotropic broadcasting has the largest scope of possible recipients.
However, the power requirements for transmitting some minimal number N(min) $\sim 10^6$ photons/sec in the 20 cm regime to a km$^2$ dish at a distance of $\sim$ 100 parsecs -10 Kiloparsec are  prohibitive, $\sim  2 \cdot 10^4 - 2 \cdot 10^6$ Gigawatt. Beaming with the maximal accuracy allowed of
$\delta(\theta) \sim \lambda/R$ with $\lambda \sim$ 20 cm and R $\sim$ Km reduces this by 10$^{-8}$!
 Note that this improvement can be achieved only if also the recipient, namely us, listen  in the same direction.}
We suggest focusing on those directions from which our solar system and the inner smaller planets, in particular, are most readily discovered.

 Four methods are presently used or are being developed for discovering planetary systems: Doppler shifts, eclipsing by transiting planets, gravitational microlensing and direct astrometric observations.

The first method using the periodic Doppler shift due to the radial velocity induced by the planet's motion may not be optimal for discovering solar-like systems. To discover our  Jupiter of mass $\sim 10^{-3}$ $M_{Sun}$ and velocity
$\sim$ 13  Km/sec, one needs to measure Doppler shifts due to a solar ``recoil" velocity of $\sim$ 13 m/sec.
The large distance from the Sun inferred from the twelve-year periodicity would suggest too cold a planet to support life. Discovering Earth, Venus or Mars via radial velocities smaller than 10 cm/sec is very difficult.

 Microlensing relies on a one-time coincidence of the lines of sight to a bright star and to a planetary system and is unlikely to serve as a basis for a systematic search.

 However, if located in the right direction the ITS could readily detect the inner stars via partial eclipsing. Earth's transit reduces the solar luminosity by 77 parts per million once a year during up to  $\sim$ 13 hours. A similar eclipsing by Venus is $\sim$ two hours shorter and reoccurs every $\sim$ seven months. A three-fold weaker eclipsing lasting up to 16 hours every 1.9 years can be induced by Mars.
Finally, mini-eclipsing by Mercury will be $\sim$ ten times weaker than by earth but will reoccur four times per year. Detecting such changes seems to be the easiest method for finding the solar system's inner planets.

 The best method is to directly observe the planets. Since earth's luminosity is $\sim 5 \cdot 10^{-10}$ that of the Sun at $5 \cdot 10^{-8}$/(R/100 parsecs) radians away for an observer at distance $R$, only nearby ITS's are likely to study Earth via this method.

 However, ITS's will be able to use the easier transit method to find Earth, if the line of sight from them to us is within an angle $\theta$(Earth) = R(Sun)/(A.U.) = 1/(200 radians) or 0.28 degrees above or below the ecliptic plane. As the earth rotates around the Sun the above lines of sight cover a ``stripe" at latitude
+/-$\theta$(Earth) of solid angle $2\pi \cdot 2\theta$(Earth)  constituting 1/200 of the sky.  Only ITS's in the ``slice" of the galaxy in the angular direction of this stripe can discover Earth via transiting.

 Similar mini solar-eclipsing by Venus could be seen by ITS's located on lines of sight at an angle $\theta$(Venus) =R(Sun)/(0.72 A.U) $\sim$ 1/(140 radians) relative to Venus's rotation plane.  These ITS's should then be in an analog slice constituting 1/140  of the sphere. The weaker Mars's eclipsings at 1.5 A.U can be seen by about 2/3 as many ITS's as compared with Earth induced eclipses. The weaker yet eclipses by Mercury at $\sim$ 0.5 A.U. will be seen at angles 0.01 radian above and below Mercury's plane.

Finally, eclipsings by Jupiter (or Saturn) with radiuses $\sim$ 10 times that of Earth last for up to a day and a half (or two), respectively, and are much stronger decreasing the solar luminosity by $\sim$ 1\%.
However, these reoccur only every twelve (or 20) years, respectively, and can be seen from stripes which are about 5 (or 10) times narrower than those for the case of eclipsing by Earth.

 The planes of planets' orbits have angles relative to the ecliptic (and to each other) exceeding
0.26$^o$  and their disjoint ``observability" slices of all planets (except Mercury) add up to $\sim$ 1.5\% of the sphere (or to 2.5\% if we add in also Mercury). We suggest focusing more on the directions of the above stripes or of the overall +/- 3.4$^o$ stripe including the individual stripes.

 The thickness of the galactic disc in our neighborhood is $\sim$ 150 parsecs. With the ecliptic at 60$^o$ relative to the disc the radial extent of the above slices where some eclipsing in the solar system is observable is typically
 $\sim$ 100 parsecs. This distance spikes at $\sim$ 10 K-parsecs towards the intersection of the ecliptic with the galactic plane.

 Thus, if we consider only those stars (and prospective ITS's thereabout) from which at any specific time eclipse by the inner planets can be seen, we restrict to 1.5\%-2.5\% of all candidates and to $\sim$ 7\% if we use the broader +/- 3.4$^o$ stripe.

 An optimistic guess of the factors in Drake's equation, assuming furthermore that ITS's survive on average a million years \cite{ChaissonMcMillan} yields about one Galactic ITS per (30 parsecs)$^3$. We then expect about 100 ``nearby"
ITS's within 100 parsecs and $\sim 10^6-10^7$ in the disc. The above discussion suggests that  only one to three
nearby ITS's can see a planet in the solar system via eclipsing at any given time.

 In the following, it is helpful to discuss separately the two classes of ``Nearby" ITS's within a 100 parsecs sphere and the many ``far" ITS's in the disc.

\vspace{.25in}

\noindent {\bf A) Nearby ITS's}

 Proper motions with velocity components---perpendicular to the line of sight and to the above stripes---of, say,
  $\sim$ 20 Km/sec cause a nearby star (at r $<$ 100 parsecs) to sweep out the full 7$^o$ stripe in less than about 600,000 years. A nearby ITS can then discover within a time shorter than its assumed average existence several solar system planets as it traverses successive ``discovery stripes" specific to individual planets.
 The potential discovery zone is further effected and in general enlarged due to nutations of the planetary planes relative to a plane fixed by $\vec{J}$(total)---close to that of Jupiter. For simplicity we will conservatively not include these in the following.

The large, $\sim$ 1\% eclipsing effect due to Saturn is unlikely to be missed.  Periodic monitoring in our direction  will later reveal the remaining planets.  Realizing that we have likely habitable inner planets protected from meteorite and astroid impacts by Jupiter, will motivate further, detailed direct observation hopefully verifying the existence of oceans and of Oxygen in our atmosphere.

 Thus, nearby ITS's in the 2.5$^o$ union of all planetary stripes or the overall  7$^o$ stripe making up 2.3\% $sim$ 6\% of all nearby ITS's may eventually beam towards us. Focusing on these narrows the search in the nearby zone by factors of 16-40.

 One can argue that very advanced ITS's can discover us from any direction via the tiny Doppler shift
 $\delta(\lambda)/{\lambda} \sim 3 \cdot 10^{-10}$  and/or by direct observations.

 Also if many ITS's existed for a long time they may have set up a communication network where only nearby  pairs of stars belonging to the net need to have direct communication. With the existence of our solar  system becoming common knowledge, the nearest ITS in the network will beam towards us.

Both of the above considerations suggest that there is no point in focusing on the special directions from where transit eclipsing can be seen. While this may be so, both arguments can be countered and a special role of the eclipsing directions is likely to remain.

 Thus, while other discovery methods may be feasible the effort required is far larger and these are more likely to be employed once the very existence of small planets has been established.  Assume then that any one or both of the radial velocities/direct observations discovery methods are available to a particular ITS.
 If further the latter is within the about 1.5\%  (or 2.5\%) of the sky from which, at any given time, the inner planets of the solar system (including or excluding Mercury) can be seen via eclipsing, they would have readily found our solar system using the transit method.  They will then {\it not} have to wait hundreds of thousands of years to have their proper motion or planetary orbit nutations to allow them to discover the remaining planets via transits.
Rather, the more advanced direct observation method which can reveal the nature of Earth atmosphere and the oceans available to them can be then be used right away to to verify the habitability of Earth.

 ITS on stars which are not in the above 2\% zone may well first find other solar systems for which these  ITS's are in the discovery stripes. These ITS's will then expend much effort studying those and if encouraged  by their findings may also eventually broadcast to them.

 The above assumed that the searches and decisions to start a communication attempt of each ITS are  made independently. We next consider the case when an extended communication network between many ITS's exists.
 Regardless of who discovered us, who will be relegated to try and communicate with us?
  We would like to argue that also in this case the choice may be the closest ITS  in the zone for our discovery via  the transit method---rather than the member of the net closest to us.  A possible rationale for this could be the following:

 The logic used in the first instance to encourage us to ``listen" to ITS's in the ``discovery-stripes" can be applied again to the would-be broadcasting ITS's. Since we are {\it not} on the galactic communication  net, we are, at best, an ``awakening technology" where the simpler transit method for discovering planetary systems is commonly used.

Realizing that we may view the above stripes as preferred directions to ``listen" in, ITS's may be encouraged to beam to us from those directions! Clearly a joint agreement to broadcast and listen in these directions enhances the effectiveness of the communication and greatly reduces the prohibitive power otherwise recurred.

(This is somewhat in the spirit of suggesting that the universal peak of the microwave---which properly redshifts  when detected at cosmologically later times---be used as a frequency standard.)

 There could be yet one additional reason why ITS's may be more likely to try communicating with us first from the above special directions.  There have been extensive discussion of the possible content of the first communication.  Using the universal language of mathematics this could be 1+1 = 2, the prime numbers or the Goldbach  conjecture that every even number equals the sum of two primes, communicated by appropriate series of pulses.
There is, however, also the option of loading into the signal shared physical knowledge such as the above microwave background peak or the following.

 Using just two properly timed pulses once every year (or once every period of another planet) the ITS can communicate the shared information of its direction relative to the ecliptic or to the plane of one the solar planets.
Thus, consider a nearby ITS which sees eclipsing of the Sun by Earth's transit repeating every year. It lasts for:
\begin{equation}
13 \, {\rm hours}  > T = 2[R(S)^2-[d_{(S \cdot E)}]^2 \, tan(\theta))^2]^{1/2}/(v_E)
\end{equation}
with $\theta$ the angle between the line of sight to the ITS and the ecliptic plane, $R(S)$ the solar radius, $d_{(S \cdot E)}$ the Earth-Sun distance ($\sim$ A.U) and $v_E \sim$ 30 Km/sec, the Earth's velocity.

   The time separation between two pulses beamed at the beginning and end of the eclipses fix T. If
  repeated every year and eventually detected by us, then the direction of the beamed signal should agree
  with the polar angle relative to the ecliptic $\theta$, inferred from T via the last equation.
  It is much more difficult for them to communicate their azimuthal location relative to us.
There is a delay of $\sim$ 300 years due to the large separation between us and the ITS. To have their signal arrive at us within the time window of another transit occurring 300 years later ``they" will have to measure the distance to the Sun with relative precision better than $10^{-6}$.

If they do achieve such a precision they can time their beaming so that it arrives here 300 years
  later at the time when they see the 300 t'h reoccurrence of the transit. We may then get a double coincidence between $\theta$ and the time
  separation T, and between the direction of the signal detected and the antipode of the Sun at this time (of night!)at the terrestrial observing station. Hopefully this direction will then point also to an actual star which is sun-like or otherwise favorable for sustaining life!
By sending pulses also in the real time of their seeing the transit they can also transmit, modulo
  a (terrestrial) light year!) the distance to their star. This distance is just the time difference
  between the series of annual pulses sent in real and shifted times.

Clearly the above applies only if the ITS wants to use this eclipsing based method.

 In passing we note that even if the Earth were known to be habitable, ITS's are unlikely to keep beaming towards it during the billions of years required for our technological civilization to emerge. However, nearby ITS's may have realized by now---or in the next 300 years---via the increased CO2 concentration, or otherwise, that technology is emerging here and would be encouraged to (re)start beaming to us.

\vspace{.25in}

\noindent{\bf  B) Far ITS's}

 Consider next the intersection of the above slices where eclipse by transiting planets is observable, and the galactic disc. If our solar system was at the galactic center with its fixed plane (about the ecliptic) perpendicular to the disc, this intersection is shaped simply as angular wedges.  The wedges of opposite pairs of ``cake-like" slices making up a fraction: $f=4/(2\pi) \cdot (\theta{\rm(Earth)} +\theta{\rm(Venus)} + \theta{\rm (Mars)} + ...$
of all the disc.  Hence 1.5\%-2\% of all stars and, in particular, of those further than three Kilo-parsecs from the center of the galaxy so as to be in the ``Galactic habitable zone" will be in this joint intersection.
The possible number of candidates using the same optimistic estimate of the fraction of stars containing ITS's is then very large, $\sim 10^5$.

The larger distances slow down in proportion to 1/R, the rate of moving between the different eclipsing stripes relative to the case of the ``nearby" ITS's. The intrinsic changes of the various planetary orbits may still enlarge the discovery scope to more than one planet by an ITS which lives longer than O (million) years.
 The number of our potential discoverers via eclipsing increase by $\sim$ 14\% by the $\beta = 60^o$ angle between the ecliptic and the disc. Also, our being about 8 Kilo-parsecs away from the center increases the average distance$^2$ to the potential discovering ITS's and places more on the away side of the noisy galactic center. All these can safely be neglected for our present purpose.

 Considering far out ITS's in the disc is particularly relevant if the optimistic estimate of the fraction of main sequence stars hosting ITS's  fails and we have only 100-1000 in the whole galaxy.

 We will then have no nearby ITS's yet O(1\%) of the above 100-1000 ITS's, namely, O(1-10), will be in the stripes of potential discovery by eclipsing.

 Consider a ``far ITS" a few kilo-parsecs away which is systematically searching for planetary systems at ever increasing distances.  Further assume that exhaustive search methods such as direct observation or radial velocities manifesting in tiny Doppler shifts are available to it. It will have to study hundreds of millions of main sequence stars before finding us. Also at increasing distances R the angular separation between the Sun and Earth  falls off as 1/R  becoming (5 $\cdot 10^{-10}$ radians for R=10 Kilo-parsecs. This is then likely to make simple eclipsing  rather than direct observation (or Doppler shifts  smaller than 10$^{-9}$) their first search method of choice. Indeed, while luminosities falls like R$^{-2}$ the $\sim 10^{-4}$ fractional drop upon Earth's transit stays the same\footnote{Clearly the distance and corresponding absolute change in intensity matters  as well. Still for the relevant galactic stars the intensity of the solar radiation may suffice.
 The sun radiates $\sim 4 \cdot 10^45$ visible photons per sec. A 10 m$^2$ telescope at a distance of  a kiloparsec will collect $\sim 4 \cdot 10^8$ photons per minute. With random fluctuations,  $\delta (N)/N \sim (1/{N})^{1/2}$,  this allows to discern a fractional change as small as $5 \cdot 10^{-5}$ and, in particular, the Earth transiting the Sun  within a minute. For an ITS which is 10 Kiloparsecs away using a 10m $\times$ 10m telescope this requires 10 minutes, $\sim \Delta$ (T), the time it takes for the minieclipsing to become maximal  or to fade away.
 The above assumed no other background, e.g.,  changing luminosity associated with a sunspot varies by $ 5\cdot 10^{-3}$\% over such a short time---an issue discussed in Sec. C below).}.

 If the far ITS which will discover ``us"---that is, Earth, Venus, Mars (or Mercury) this
way---will beam to us, the signal will be from Earth's, Venus's, etc., eclipsing stripes.

 We note that in general the number of ITS's in the disc which can potentially discover any given planetary system with an invariant plane inclined at an angle $\beta$ relative to the disc grows as $1/(sin(\beta))$ saturating at
 $\beta \sim 0.01$ the aspect ratio of our disc. Thus planetary systems aligned with the disc are exposed via mini-eclipsing to more observers and are likely to be discovered first.

\vspace{.25in}

\noindent{\bf  C) Possible Special (Local) Obstacles for Our Discovery via the Transit Method}

 The transit method successfully revealed many exo-solar planets. Unfortunately special features of our  solar system and Sun may hinder the efforts of those ITS's situated in the discovery angular patches to discover us via this method.
  Our Sun may be varying thereby obscuring the desired small yearly signal due to Earth's transit.
The total energy output of the ``central fusion engine" at the solar core remains constant during the
$\Delta(T) \sim$ 1/4 hour required for Earth's shadow to enter into or re-emerge from the solar disc.
Indeed the latter time may be also shorter than the overall time of solar quakes.
{Even putative variations of the central engine on  scales of, say, thousands of years are quenched  by the long time required for the photons to diffuse from the core to the solar surface.

  The most severe likely difficulty is the ``noisy" solar surface and related variations of solar luminosities. These could potentially reflect (dis)appearance of sunspots which can die  away (or form) and move to (emerge from) from the opposite side of the Sun. The spots may be rather  big, even of Earth size, and colder by $\sim$ 15\%- 20\% than the average solar surface temperature. Can  they mimic the T $\sim$ O(10) hour slight (60 ppm) dimming which turns on and off rather sharply (within $\Delta (T) \sim$ 15 minutes) due to an Earth-like planet transiting in front of the sun?

We believe that this is not the case. The spots are a local surface feature in part due to locally  enhanced magnetic fields. Such fields deflect and channel  some of plasma/heat flow away from  the sunspots to nearby regions---generally on the same side of the Sun.  Indeed sunspots indicate a more  active and potentially more luminous sun than otherwise.  All this and the longer time scales for the evolution of sunspots and associated prominence and arcs make mimicking of the transit effects unlikely.

To decide this issue we need to measure solar luminosity variations on time scales shorter than
 $\Delta(T) \sim$ 15 minutes.  The best outcome from the point of view of making our suggestion
sensible is that no or very few such short-term variations of order 50 ppm occur in a year. More likely this will predominantly be the case during the period of a ``quiet sun" with few or no sunspots.

The prospective Kepler satellite is believed to be able to detect an Earth-like planet on the background of a noisy solar-like surface.  This should be all the more the case for the prospective much more advanced ITS!

 The solar luminosity has been monitored over the last 30 years by  satellites. The main mission was to study solar cycles and longer trends. Thus minute-by-minute readings  are not available in the public domain.
 If carried over hundreds of years these measurements will find many opportune time windows when no or few  spots are present and the Sun is more quiescent and stable. During such periods (shifted by the light  travel time (LTT) which may be as large as LTT $\sim$ 300 years), it will be easier for the ITS's to discover us via the  transit method.  It is amusing to see how this may help if the ITS would transmit as we suggested above at  the time when they see the transit and a time shifted by LTT. We can  check if a series with less pulses  corresponding to actual observations corresponds to times when our Sun was quiescent.

 2) The astroid belt might interfere with the transit eclipse measurements by providing a background of many micro-eclipsings. This, however, is not the case. The belt constitutes a well-defined narrow disk which does (partially) eclipse the Sun when viewed from an appropriate ``patch". However, this patch is distinct from the other patches where eclipses by the other planets can be viewed. Further, unlike the periodic eclipsing by the planets that is due to the astroid belt is essentially constant.

 3) An interesting  ``complication" stems from perturbations due to the moon. With[R(moon)/R(earth)]$^2 \sim$ 0.07 it will induce additional doubly periodic smaller eclipses and tiny modulations of order $3 \cdot 10^{-4}$ of Earth's transit time. These effects are too small to mask the primary eclipsing by Earth. Yet they can indicate the existence of a moon and provide more data fixing better the planet's mass and density.

\vspace{.25in}

\noindent{\bf  D) Conclusions}

We suggested that certain directions in the sky---the above patches where solar eclipses by transiting planets are viewable---may be preferred search directions and that radio signals coming therefrom should be more carefully analyzed.  Since we are more readily discovered by ITS's in these directions, we may be more likely approached by radio signals from these directions as well.

 We have attempted to address obvious caveats to this apparently simplistic argument for both nearby and far ITS's which are, in particular, relevant for  many, say, more than a million ITS's in the galaxy and relatively few, say, 100-1000 ITS's, respectively.

  We have clearly emphasized further difficulties associated mainly with the noisy solar surface  which can impede detection, but definitely will not prevent it. On the more pessimistic side, we note that  if transit was the primary method of discovering potential habitable stars utilized by ITS's, then this may  help answer the famous Fermi question: ``Where are They?". While they visited many potentially habitable  planets that they discovered they simply have not discovered us. This is due to the confluence of 1) our ecliptic plane being inclined by 60\% to that of the galaxy---hiding us from most potential ITS's which are actively searching; and 2) a noisy Sun surface further impeding discovery via the transit method.

 We realize that ``second-guessing" aliens, which was actually attempted here, is highly speculative.  Also if there  are altogether less than $\sim$ 10-50 ITS's there may be none in the above patches and our suggestion will  not help the SETI or other future listening project (which will be then facing huge difficulties anyway).  All that notwithstanding we believe that our suggestion has some merit and deserves further study.

\vspace{.25in}

\noindent{\bf Acknowledgements}

  Being a complete novice to this field, I have greatly benefitted from many helpful comments,  encouragements and criticism. I am indebted to P. Goldreich, J. Ostriker, and D. Spergel for  increasing degrees of the latter; to the young A. Cowsick for pointing the importance of sunspots;  to A. Botero for correcting an embarrassing error in an earlier version and to P. Steinhardt  for emphasizing the effects of proper motion. I have enjoyed helpful discussions  with  Y. Aharonov,  R. Cowsik, F. Dyson, T. Jacobson, J. Goodman, J. Kanner, A. Laor, T. Nussinov Z. Nussinov, R. Shrock,  J.Tollaksen and I. Wasserman.
 I am particularly indebted to T. Mazeh from whom I learned about exo-solar planets early on; to D. Kazanas and to A.  Smith for many conversations and considerable help; and foremost to S. Tremaine for continuous encouragement and very useful information re the solar system and the very pertinent work of J. R. Gott.

\end{document}